\documentclass[fleqn,twoside]{article}
\usepackage{espcrc2}
\usepackage{graphicx}

\newcommand {\beq} {\begin{eqnarray}}
\newcommand {\eeq} {\end{eqnarray}}
\newcommand {\eol} {\nonumber \\}

\title
{\Large A study of colour field distributions in the baryon}

\author{F. Okiharu\address{Department of Physics, 
        Faculty of Science and Technology, Nihon University, \\
        1-8-14, Kanda-Surugadai, Chiyoda-ku, 
        Tokyo, 1018308, Japan} 
        and 
        R. M. Woloshyn\address{TRIUMF, 4004 Wesbrook Mall,
        Vancouver, BC, V6T 2A3, Canada}}

\begin{document}

\begin{abstract}
The distributions of chromo-electric and chromo-magnetic field 
associated with flux tubes in the baryon are studied in SU(3) lattice QCD. 
Maximal Abelian projection is used to reduce 
the statistical fluctuations. 
For a fixed source geometry, many different string configurations 
are possible. We investigated whether the string configuration, 
that is the choice of operator, 
biases the observed flux distribution.
\vspace{1pc}
\end{abstract}

\maketitle

\section{Introduction}

Recently, the problem of how chromo-electric and chromo-magnetic 
fields are distributed in a baryon has been of considerable interest. 
It has been investigated in terms of the potential \cite{Taka01,Alex02}.
But the flux distribution can be calculated directly using lattice QCD. 
There are many studies for a meson \cite{Somm86,Bali95,Haym96}, 
but only a few for a baryon. 
The first attempt to do this was by Flower\cite{Flow86}
but it was very noisy and not conclusive. 
However, in the past year, Ichie {\it et al.}\cite{Ichi02}
have presented very nice results using Abelian projected fields.

Generally speaking, all lattice simulations have some systematic errors 
and it is important to have some estimate of these errors 
in order to assess the results. 
What are the sources of systematic error in the map of flux distribution? 

One is interested in the shape of the flux distribution. However, 
in order to do the calculation some choice for the three quark operator 
has to be made. This operator consists of three quarks connected 
by strings of gauge field links which assume some shape. 
The systematic effect which we study here is 
whether the shape of the operator influences the measured flux distribution. 
 
Of course, the ground state property in a lattice simulation should be 
independent of the operator. But the conditions needed to obtain 
this ideal situation  may be different for different quantities 
and may be difficult to achieve in practice. 
This can be seen by considering the time dependence of the two-point and 
three-point correlation functions. 

For the two-point function \\
\beq
& & \langle \Omega |O(T)O(0)|\Omega \rangle \eol
&=& \sum_n e^{-E_nT}
\langle \Omega|O(0)|n\rangle \langle n |O(0)|\Omega \rangle \eol
& \to & e^{-E_0T}
\langle \Omega|O(0)|0 \rangle \langle 0 |O(0)|\Omega \rangle .
\eeq

For the three-point function ($T>t$) \\
\beq
& & \langle \Omega |O(T)P(t)O(0)| \Omega \rangle \eol
&=& \sum_{n,n'}e^{-E_n(T-t)}e^{-E_{n'}t} \eol
& & \times \langle \Omega|O(0)|n \rangle 
\langle n|P(0)|n' \rangle 
\langle n' |O(0)| \Omega \rangle \eol
&\rightarrow& e^{-E_0T}
\langle \Omega|O(0)| 0 \rangle 
\langle 0 |P(0)|0 \rangle 
\langle 0 |O(0)| \Omega \rangle . 
\eeq

The generic form of the correlation function which gives the field
distribution is \\
\beq
& & \frac{\langle \Omega |O(T)P(t)O(0)| \Omega \rangle}
{\langle \Omega |O(T)O(0)| \Omega \rangle} 
- \langle \Omega |P(0)| \Omega \rangle \eol
&\rightarrow& \langle 0 |P(0)| 0 \rangle - 
\langle \Omega |P(0)| \Omega \rangle .
\eeq

In this case, the condition $T>>t>>0$ is required 
to isolate the ground state contribution.
For the two point function, which yields the potential, 
one only needs $T>>0$. If a simulation shows a dependence 
on the choice of operator this indicates that there are 
non-ground state contributions.

\section{Simulation}

We studied the three quark potential and the field distributions. 
The three quark Wilson loop is defined as 
\beq
W_{3Q}=\frac{1}{3!} \epsilon_{abc}\epsilon_{a'b'c'} 
       U^{aa'}_1 U^{bb'}_2 U^{cc'}_3
\eeq
with the path-ordered link variables
\beq
U_j \equiv 
P \exp \Bigl\{ig \int_{\Gamma_j} dx_\mu A^\mu(x) \Bigr\} \quad (j=1,2,3).
\eeq
The path is denoted by $\Gamma_j$ in Fig. 1.

\begin{figure}[t]
\includegraphics[width=53mm]{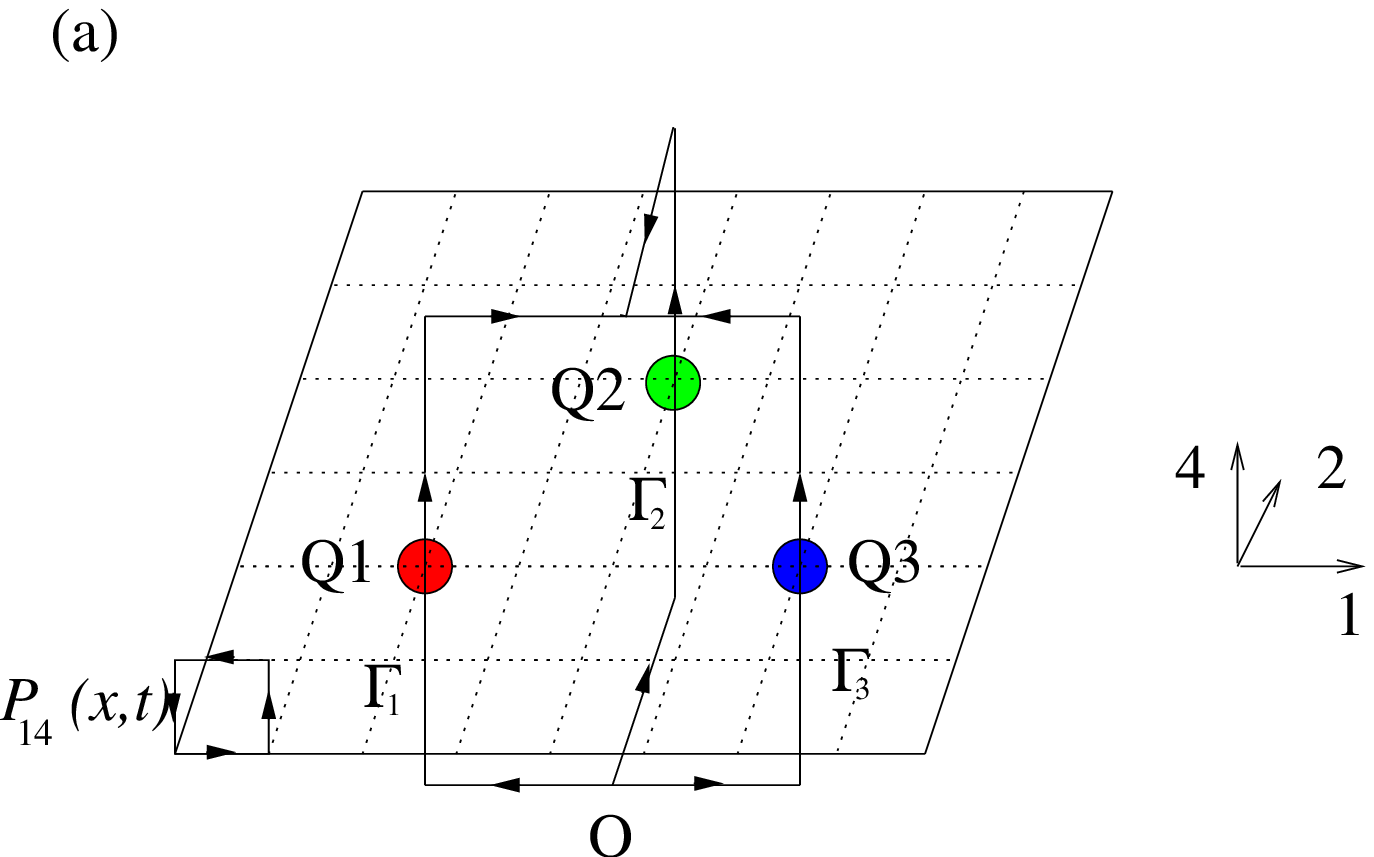}
\includegraphics[width=53mm]{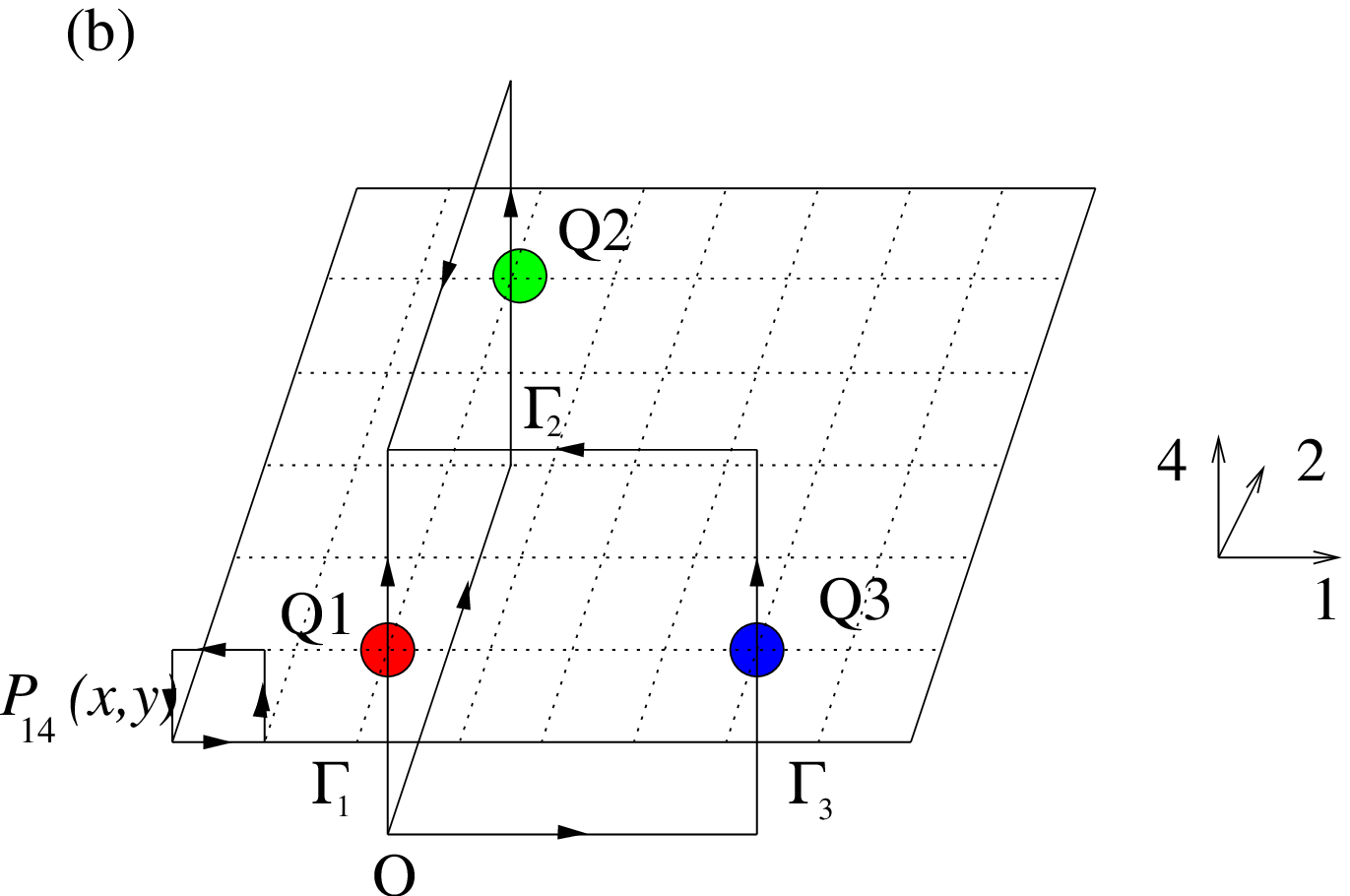}
\caption{The three quark Wilson loop (a) T-shape string configuration
 (b)L-shape string configuration.}
\end{figure}

The correlation function can be described as \\
\beq
P_{\mu \nu}(x,t)
=\frac{\langle P_{\mu \nu}(x,t)W_{3Q}(T) \rangle}{\langle W_{3Q}(T) \rangle} 
- \langle P_{\mu \nu}(t) \rangle.
\eeq
Here, $P_{\mu \nu}$ denotes an unsmeared plaquette, 
used as a probe of the chromo-electric or chromo-magnetic field. 
In our simulation the plaquette insertions were made at $t=T/2$.

The simulation is done in quenched QCD 
using the Wilson plaquette action with $\beta = 6.2$ on $24^4$ lattices. 
In order to be able to study systematic effects 
it is necessary to control statistical errors.
The multi-hit procedure was used for temporal links in order to reduce the 
statistical fluctuations in the potential calculation.
In addition, smearing of spatial links is used to enhance the ground state.
The flux distribution calculation is very noisy. 
Ichie {\it et al.}\cite{Ichi02} have shown that maximally Abelian projection, 
which preserves the correct long distance behavior of the Wilson loop, 
can be used very effectively to investigate the flux tube in the baryon. 
We adopt this method here. 

Fig. 1 shows the choice of the operator. 
Three quarks are located on the $x-y$ plane 
with the distance $k$ from the origin, 
$(Q1(x,y),Q2,Q3)=((-k,0),(0,k),(k,0))$ in (a), 
$((0,0),(0,k),(k,0))$ in (b). 
The operator (b) has not been considered before. 

\section{Results and Discussion}

\begin{figure}[h]
\includegraphics[width=53mm,angle=90]{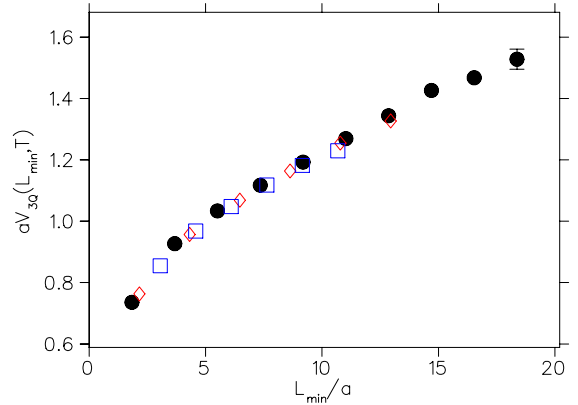}
\caption{The three quark potential.}
\end{figure}
Fig. 2 shows the three quark potential for different operators. 
$L_{min}$ describes the total minimum length from the physical junction 
to each quark. 
The circles denote the case of three quarks located on the
axes at the same distance $k$ from the origin, 
while the triangles and the squares show the quark geometries of 
Fig.1 (a) and (b) respectively. 
There is no evidence for any operator dependence. 
\begin{figure}[h]
\includegraphics[width=53mm]{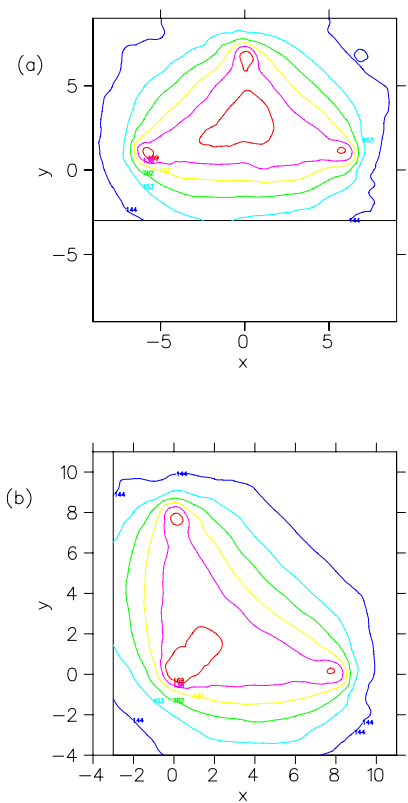}
\caption{The action density (a) T-shape string configuration 
(b) L-shape string configuration.}
\end{figure}

Fig. 3 shows the action density for operators with different
string configurations. 
The three quark positions, $((-6,0),(0,6),(6,0))$ in (a) 
and $((8,0),(0,0),(0,8))$ in (b) are chosen 
since the inter-quark separations are similar. 
$T=8$ is considered since it is large enough to extract the ground state 
for the potential. 
There a visible difference between the shapes of the distributions 
which are still biased by the form of the operator.
This operator dependence suggests that excited states are still contributing.

Ideally, the physical ground state property in a lattice simulation 
should be independent of the choice of the operator. 
For the three quark potential, this seems to be readily achievable. 
However, since the flux distribution contains the three-point function,
it requires large time separation not just between the quark source and sink 
but also between the source and sink and the flux probe.
These conditions are not easy to meet 
and in our simulations, operator dependence in the flux distribution is visible. 
It is important for future calculations to check for this systematic 
effect, especially if it is claimed that the measured flux distribution 
has the same shape as the strings contained in the baryonic operator.

To reduce the systematic error discussed here, 
even more effective techniques to reduce statistical fluctuations 
would be helpful. 
This would allow calculations at larger time separations. 
Alternatively, operators without strings, as used, 
for example, by the MILC Collaboration\cite{MILC} to calculate 
the quark-antiquark potential, might be useful.

This work was supported in part by the Natural Sciences and Engineering 
Research Council of Canada.

\end{document}